\documentclass[aps,prd,twocolumn,superscriptaddress,nofootinbib]{revtex4-2}
\usepackage[utf8]{inputenc}
\usepackage[T1]{fontenc}
\usepackage[english]{babel}
\usepackage{ragged2e}
\usepackage{amsmath}
\usepackage{graphicx} 
\usepackage{physics} 
\usepackage{braket} 
\usepackage{times}
\usepackage{hyperref}
\hypersetup{colorlinks=true,linkcolor=blue,citecolor=cyan}
\usepackage{float}
\usepackage{amssymb}
\usepackage{orcidlink}

\def\ii{\text{i}}
\def\ee{\operatorname{e}}

\begin{document}
\title{The relativistic reason for quantum probability amplitudes}

\author{Karol Sajnok\orcidlink{0009-0004-5899-8923}}
\email{ksajnok@cft.edu.pl}
\affiliation{Institute of Theoretical Physics, University of Warsaw, Pasteura 5, 02-093 Warsaw, Poland}
\affiliation{Center for Theoretical Physics, Polish Academy of Sciences, Aleja Lotników 32/46, 02-668 Warsaw, Poland}

\author{Kacper Dębski\orcidlink{0000-0002-8865-9066}}
\email{kacper.debski@upol.cz}
\affiliation{Department of Optics, Palacký University, 17. listopadu 1192/12, 771 46 Olomouc, Czech Republic}
\affiliation{Institute of Theoretical Physics, University of Warsaw, Pasteura 5, 02-093 Warsaw, Poland}

\author{Andrzej Dragan\orcidlink{0000-0002-5254-710X}}
\email{dragan@fuw.edu.pl}
\affiliation{Institute of Theoretical Physics, University of Warsaw, Pasteura 5, 02-093 Warsaw, Poland}
\affiliation{Centre for Quantum Technologies, National University of Singapore, 3 Science Drive 2, 117543 Singapore, Singapore}

\date{\today}

\begin{abstract}
We show that the quantum-mechanical probability distribution involving complex probability amplitudes can be derived from three natural conditions imposed on a relativistically invariant probability function describing the motion of a particle that can take multiple paths simultaneously. The conditions are: (i) pairwise Kolmogorov additivity, (ii) time symmetry, and (iii) Bayes' rule. The resulting solution, parameterized by a single constant, is the squared modulus of a sum of complex exponentials of the relativistic action, thereby recovering the Feynman path-integral formulation of quantum mechanics.
\end{abstract}

\maketitle

\section{Introduction}

In his celebrated \emph{Lectures on Physics}, Richard Feynman argued that the essential principles of quantum theory can be understood by analyzing the iconic Young double-slit experiment. In this experiment, a single particle appears to traverse both slits simultaneously. The outcome of each individual trial is fundamentally unpredictable, and computing the probability of a given result requires a formalism based on complex probability amplitudes. ``How does it work? What is the machinery behind the law?’’ Feynman asked. ``We have no ideas about a more basic mechanism from which these results can be deduced’’, he concluded~\cite{Feynman1965LecturesIII}.

Dragan and Ekert~\cite{DraganEkert2020}, along with subsequent extensions of their framework~\cite{Dragan2022}, proposed that the peculiar behavior of a single quantum particle can be inferred from a more fundamental physical principle: an extension of the relativity postulate to superluminal frames of reference (see also the discussion in Refs.~\cite{grudka2022comment, Dragan2022reply, del2023comment, dragan2023reply2, horodecki2023comment, Dragan2023reply, gavassino2022fundamental, grudka2023superluminal, paczos2024covariant, jodlowski2024covariant, lake2025, sen2025superluminal, Damski2025, michalski2025stories}). They argued that the emergence of complex probability amplitudes can be viewed as a direct consequence of the relativistic invariance of probability distributions, but provided only a sketch of the argument rather than a full derivation.

In this work, we fill in this gap and show that quantum theory follows from much simpler ground rules. We demonstrate that the simplest nonclassical theory that is relativistically invariant and satisfies canonical probability axioms is quantum theory. The rule of summing complex amplitudes over paths and then squaring is not an arbitrary postulate, but the simplest solution consistent with relativistic invariance and probabilistic composition \cite{bayes}, formulated within a minimally nonclassical extension of Kolmogorov probability theory \cite{Kolmogorov1950}. In particular, we prove a closed uniqueness result: the only bounded, nontrivial family compatible with our three assumptions is the squared modulus of a sum of complex exponentials, i.e., the Feynman path-integral prescription \cite{feynman1948space,FeynmanHibbs1965}.

Early attempts to account for the mystery of quantum mechanics begin either with the assumption of a complex Hilbert space \cite{von2018mathematical,mackey2004mathematical,gleason1975measures} or with its lattice-theoretic (logical) proxy \cite{birkhoff1975logic,piron1976foundations,soler1995characterization}. In the Hilbert-first line, the Hilbert space and operator calculus are taken as primitive: observables are represented by self-adjoint operators and events by projections, while states define $\sigma$-additive probability measures on these events, with the Born rule (and its trace form) emerging as the natural/essentially unique assignment once this structure is assumed \cite{von2018mathematical,mackey2004mathematical,gleason1975measures}. The lattice-logical programme instead axiomatizes the event structure itself as an orthomodular lattice of propositions; representation theorems then recover Hilbert-type geometry from lattice properties, with Sol\`er showing that under a mild richness condition the underlying scalar field is restricted to $\mathbb{R}$, $\mathbb{C}$, or $\mathbb{H}$ \cite{birkhoff1975logic,piron1976foundations,soler1995characterization}. While these approaches constrain the kinematics of states and events, they do not by themselves explain why probabilities are computed via the ``add amplitudes, then square'' rule. A complementary perspective is provided by Sorkin's hierarchy of interference \cite{sorkin1994quantum,sorkin2007quantum}, which organizes admissible deviations from classical Kolmogorov additivity. Quantum theory is characterized by the absence of irreducible higher-than-second-order interference, meaning that any three- (and hence $n$-) alternative pattern is fully determined by the corresponding one- and two-alternative configurations; this prediction has been tested in multi-slit interferometry with no evidence for higher-order terms within present bounds \cite{sinha2010ruling,kauten2017obtaining}. A later line of research starts from general probabilistic consistency: Cox's analysis derives the sum/product rules and Bayes' theorem as the universal chain rule for coherent conditional probabilities \cite{cox1946probability,cox2001algebra}. Modern operational reconstructions embed this Bayesian syntax into frameworks where experiments compose sequentially and in parallel, and where causality/no-signalling-from-the-future ensures well-defined marginals and conditioning; quantum theory is then selected by further informational or structural axioms (e.g.\ purification, tomography, information constraints) \cite{Wheeler1989,hardy2001quantum,hardy2011reformulating,fuchs2002quantum,zeilinger1999foundational,brukner2003information,brukner2009information,grinbaum2003elements,chiribella2011informational,chiribella2016quantum,wootters1981statistical,BohrUlfbeck95,rovelli1996relational,clifton2003characterizing,DAriano08,masanes2011derivation,muller2012structure}. Related works on no-signalling theories beyond quantum correlations, symbolized by PR boxes, show that relativistic no-signalling alone is too weak and motivates additional principles that carve out the quantum set from the larger no-signalling polytope \cite{popescu1994quantum,dam2005,brassard2006limit,barnum2007generalized,pawlowski2009information,allcock2009recovering,navascues2015almost}. Finally, Goyal and collaborators develop a reconstruction that targets the amplitude algebra itself \cite{goyal2008,goyal2010origin,knuth2012foundations,goyal2014}. Their key move is to assign to each experimentally defined sequence an ordered pair of real numbers and demand consistency of \emph{serial} (concatenation) and \emph{parallel} (coarse-graining) composition rules. Under associativity/distributivity-type constraints and mild regularity, these pairs are forced to combine exactly as complex numbers, reproducing Feynman's sum and product rules and the modulus-squared probability rule; additional operational input can then be used to recover the finite-dimensional Hilbert-space formalism. In this sense, their programme derives complex amplitudes from compositional consistency, but without tying the composition law to relativistic path invariants.

Our objective is to demonstrate that none of the aforementioned sophisticated approaches are required to achieve a fundamental understanding of quantum theory, and that the probabilistic, quantum-mechanical description of dynamics in fact constitutes the simplest non-classical and covariant framework available. The emergence of non-deterministic dynamics and the presence of quantum superpositions were previously shown in Ref.~\cite{DraganEkert2020} to arise as consequences of the Galilean principle of relativity. In the present work, we complete this line of reasoning by demonstrating how the formalism of complex probability amplitudes follows from equally foundational assumptions.

The paper is organized as follows. In Sec.~II we formulate the problem by introducing three physical requirements: pairwise Kolmogorov additivity, time symmetry, and Bayesian composition. In Sec.~III we derive the general form of all probability functions that satisfy these conditions. Concluding remarks are given in Sec.~IV.

\section{Claim}

We show that three natural conditions --- \eqref{eq:pairwise-decomposition} Kolmogorov additivity restricted to one- and two-path contributions (no irreducible higher-order interference), \eqref{eq:time-symmetry} time symmetry, and \eqref{eq:bayes-rule} Bayesian composition --- imposed on a relativistically invariant family of probability functions $\mathcal{P}^n$ uniquely reconstruct the quantum-mechanical probability distribution.

\begin{figure}
    \centering
    \includegraphics[width=0.48\textwidth]{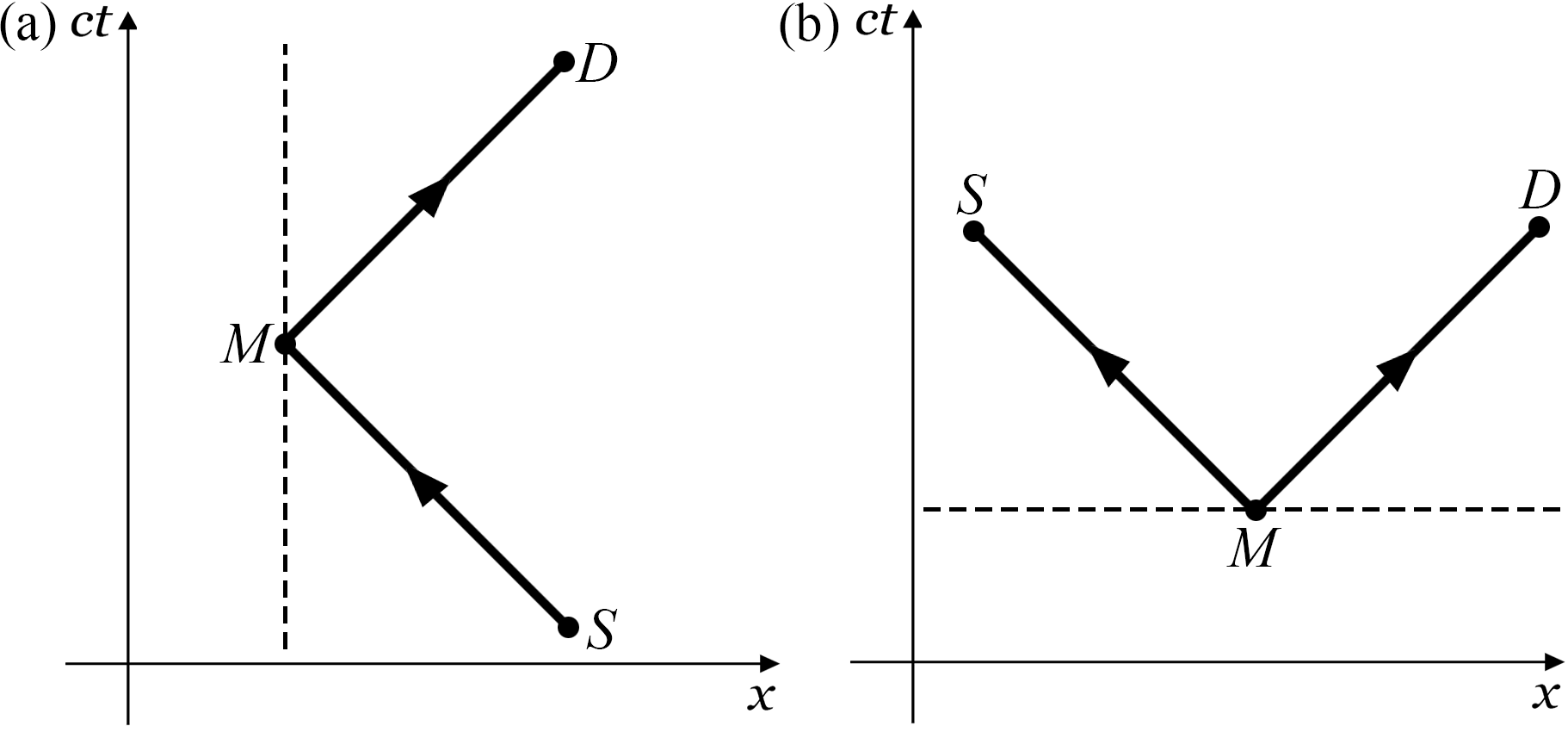}
    \caption{\justifying Mirror thought experiment illustrating the relativistic reinterpretation of a single broken worldline. (a) In a standard subluminal inertial frame (coordinates $ct,x$), a particle emitted at $S$ is reflected at the mirror event $M$ and detected at $D$, forming a single continuous trajectory $S\to M \to D$. (b) In the superluminal-branch description (axes interchanged), the same spacetime process is represented as two concurrent legs emanating from the reflection event $M$, connecting to $S$ and $D$. The dashed line indicates the mirror worldline. This illustrates how a description based on a single classical path becomes frame-dependent once the superluminal branch is admitted, motivating a probabilistic assignment to sets of alternative trajectories. Adapted from \cite{DraganEkert2020}.}
    \label{fig:mirror}
\end{figure}

The relativistic motivation for adopting a multi-path description is made explicit in the mirror thought experiment of Dragan and Ekert \cite{DraganEkert2020}, illustrated in Fig.~\ref{fig:mirror}. Consider a photon emitted at spacetime point $A$, reflected at a mirror located at $M$, and subsequently detected at $B$. In a standard subluminal inertial frame, this process is naturally represented by a single broken worldline $A \to M \to B$. By contrast, in a superluminal branch frame the same spacetime process is described in terms of two concurrent segments emerging from the reflection event, namely $M \to A$ and $M \to B$. If which-path detectors are introduced along these segments, at most one detector can register a click, but in the absence of such measurements the superluminal description necessarily involves a coherent superposition of both alternatives. Consequently, the assumption of a single classical worldline that remains valid in all inertial frames becomes incompatible with an extension of relativistic invariance to the superluminal sector. This incompatibility motivates a reformulation in terms of sets of multiple trajectories rather than a unique path. Dragan and Ekert \cite{DraganEkert2020} further show how this reasoning extends to scenarios involving an arbitrary number of possible paths.

We begin with the standard Kolmogorov probability measure \cite{Kolmogorov1950} on an event algebra. For two events $A,B$ we have
\begin{align} \label{eq:kolmogorov2}
    P(A \cup B) = P(A) + P(B) - P(A \cap B),
\end{align}
where the overlap term $P(A \cap B)$ encodes the joint occurrence of $A$ and $B$. For three events $A,B,C$, the inclusion--exclusion principle gives
\begin{align}
P(A \cup B \cup C)
&= P(A) + P(B) + P(C) \nonumber\\
&\quad- P(A \cap B) - P(A \cap C) - P(B \cap C) \nonumber \\
&\quad+ P(A \cap B \cap C).
\end{align}

\begin{figure*}
    \centering
    \includegraphics[width=\textwidth]{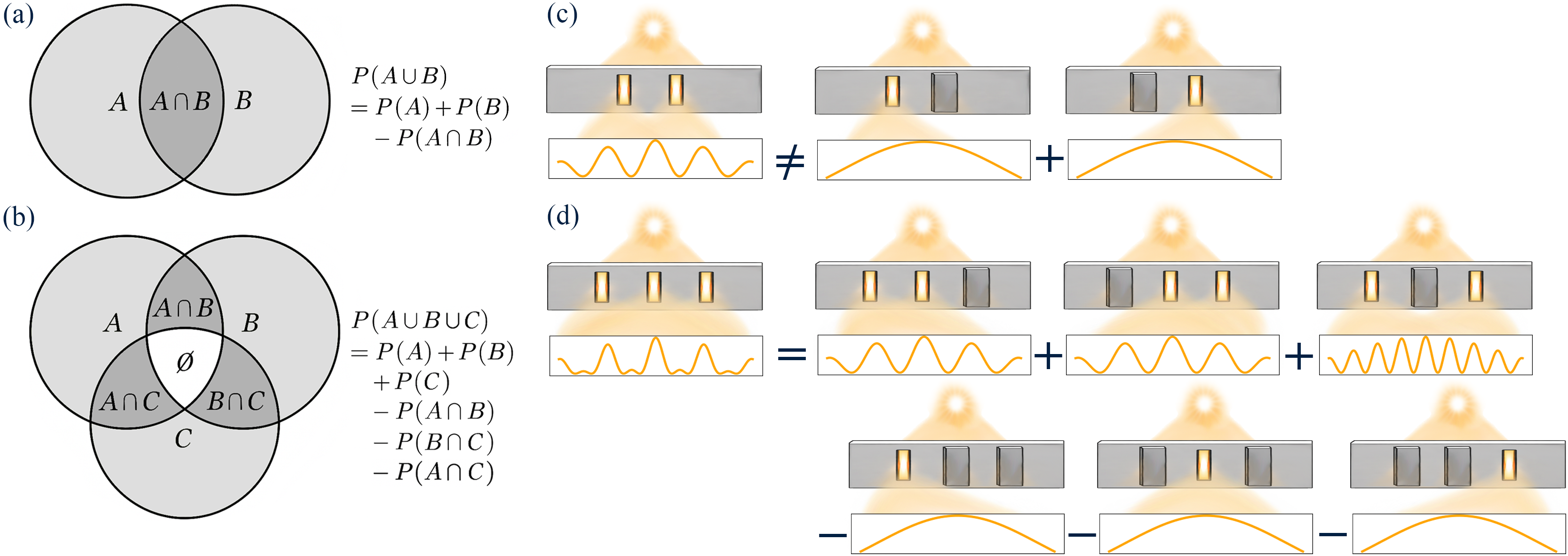}
    \caption{\justifying Illustration of pairwise Kolmogorov additivity (equivalently, the absence of irreducible higher-than-second-order interference). Panels (a,b) provide a probability-space schematic: (a) for two events $A$, $B$ the union contains a pairwise overlap term, while (b) for three events $A$, $B$, $C$ all genuine three-way overlaps are set to zero, so the union is determined entirely by one- and two-event contributions. Panels (c,d) illustrate the same principle for multi-slit interference in the far-field (Fraunhofer) regime. In the double-slit case (c), the two-slit intensity is not the sum of the two single-slit patterns because of a nontrivial pairwise interference term. In the triple-slit case (d), by contrast, the absence of genuine third-order interference implies closure at the pairwise level: the three-slit pattern is completely fixed by the collection of single-slit (two slits closed) and double-slit (one slit closed) intensity profiles.}
    \label{fig:kolmogorov}
\end{figure*}

In interference experiments with multiple slits we typically deal with mutually exclusive alternatives, such as ``the particle went through slit $i$’’ or ``slit $j$’’. Classically, these events cannot occur simultaneously, so for $i\neq j$ one has $P(A_i \cap A_j)=0$, and the probability of the union of such alternatives is simply the sum of the single-event probabilities. In a quantum theory, however, we allow for effective pairwise ``overlaps’’ even between such mutually exclusive alternatives, interpreting them as interference terms — a notion that can physically emerge even for classically exclusive events when relativistic invariance is extended to superluminal observers \cite{DraganEkert2020}. In superluminal frames, classically exclusive trajectories may jointly contribute to the physical description, as if a particle followed several paths simultaneously, allowing $P(A_i \cap A_j)$ to be interpreted as a relativistically induced form of nonclassical behavior. The simplest departure from classicality is obtained by allowing \emph{pairwise} intersections to be nonzero while requiring all \emph{higher-order} intersections to vanish. This axiom is illustrated by the Venn diagrams in Fig.~\ref{fig:kolmogorov}~(a,~b).

If the higher-order intersection terms vanish, i.e.\ $P(A_{1} \cap A_{2} \cap \ldots \cap A_{i}) = 0$ for all $i \ge 3$, we can use the inclusion--exclusion principle to compute the probability of the union of $n$ events $\{A_i\}$:
\begin{align}
    P\left(\bigcup_{i=1}^n A_i\right)
    = \sum_{i=1}^n P(A_i)
    - \sum_{1 \le i < j \le n} P(A_i \cap A_j).
\end{align}
Finally, using Eq.~\eqref{eq:kolmogorov2} to express the intersections in terms of unions, we obtain
\begin{align} \label{eq:sorkin}
    P\left(\bigcup_{i=1}^n A_i\right)
    = \sum_{1 \le i < j \le n} P(A_i \cup A_j)
    - (n-2) \sum_{i=1}^n P(A_i).
\end{align}
Equation~\eqref{eq:sorkin} states that the probability of observing any of the $n$ alternatives is fully determined by the probabilities of all single-alternative and all two-alternative configurations. No irreducible higher-order contribution is allowed. This is analogous to the Sorkin's condition of ``no higher-order interference'' \cite{sorkin1994quantum,sorkin2007quantum}, expressed purely in terms of observable probabilities rather than abstract interference functions.

To remain as close as possible to this Kolmogorov picture while allowing for nonclassical behavior, we now translate Eq.~\eqref{eq:sorkin} into the language of relativistically invariant probability functions \cite{DraganEkert2020}. We consider a set of $n$ alternative spacetime trajectories  $x_i(t)$, $i \in \{1,\ldots,n\}$, which represent the various possible paths that could be taken by a quantum particle. We denote by $\mathcal{P}^n(\{\Phi_i\})\equiv \mathcal{P}^n(\Phi_1,\Phi_2,\ldots,\Phi_n)$ the probability density of transition from the source $S$ to the detector $D$ through $n$ such paths, where $\Phi_i$ is some relativistic invariant so that the whole expression is also relativistically invariant. The simplest choice of $\Phi_i$ is the proper time along the path, which (up to a multiplicative constant) can be written either as the integral of the Lorentz factor or, equivalently, in terms of the instantaneous energy $E_i(t)$ and momentum $p_i(t)$:
\begin{align}
\Phi_i &\propto \int\limits_S^D \sqrt{1 - \dot{x}_i^2(t)}\,\mathrm{d}t \propto \int\limits_S^D \bigl(E_i(t)\,\mathrm{d}t - p_i(t)\,\mathrm{d}x\bigr) = -S_i,
\end{align}
where $S_i$ is the action of the particle along the path $x_i$. While the proper time is not the only possible relativistic invariant along a trajectory, it is the simplest one, depending solely on the instantaneous velocity rather than on higher-order derivatives of the path \cite{Dragan2021}. 

\emph{Pairwise Kolmogorov additivity.}  
In the language of the probabilistic functions $\mathcal{P}^n(\{\Phi_i\})$, Eq.~\eqref{eq:sorkin}, for $n\ge 2$, becomes
\begin{align} \label{eq:pairwise-decomposition}
    \mathcal{P}^n(\{\Phi_i\})
    = \sum_{1 \le i < j \le n} \mathcal{P}^2(\Phi_i,\Phi_j)
    - (n-2)\sum_{i=1}^n \mathcal{P}^1(\Phi_i). \tag{$\star$}
\end{align}
In words: the detection intensity for $n$ alternatives is completely determined by the one-path intensities and the two-path contributions. No additional, irreducible $i$-path terms with $i\ge3$ are present. This statement --- asserting that all multi-slit patterns are determined by their single- and double-slit configurations --- is physically motivated by the three-slit interference experiments illustrated in Figure~\ref{fig:kolmogorov}~(c,~d).

Let $\mathcal{P}^n$ be a smooth function of $n$ invariants and, in addition to the pairwise decomposition~\eqref{eq:pairwise-decomposition}, impose the following two conditions.

\begin{figure*}
    \centering
    \includegraphics[width=\textwidth]{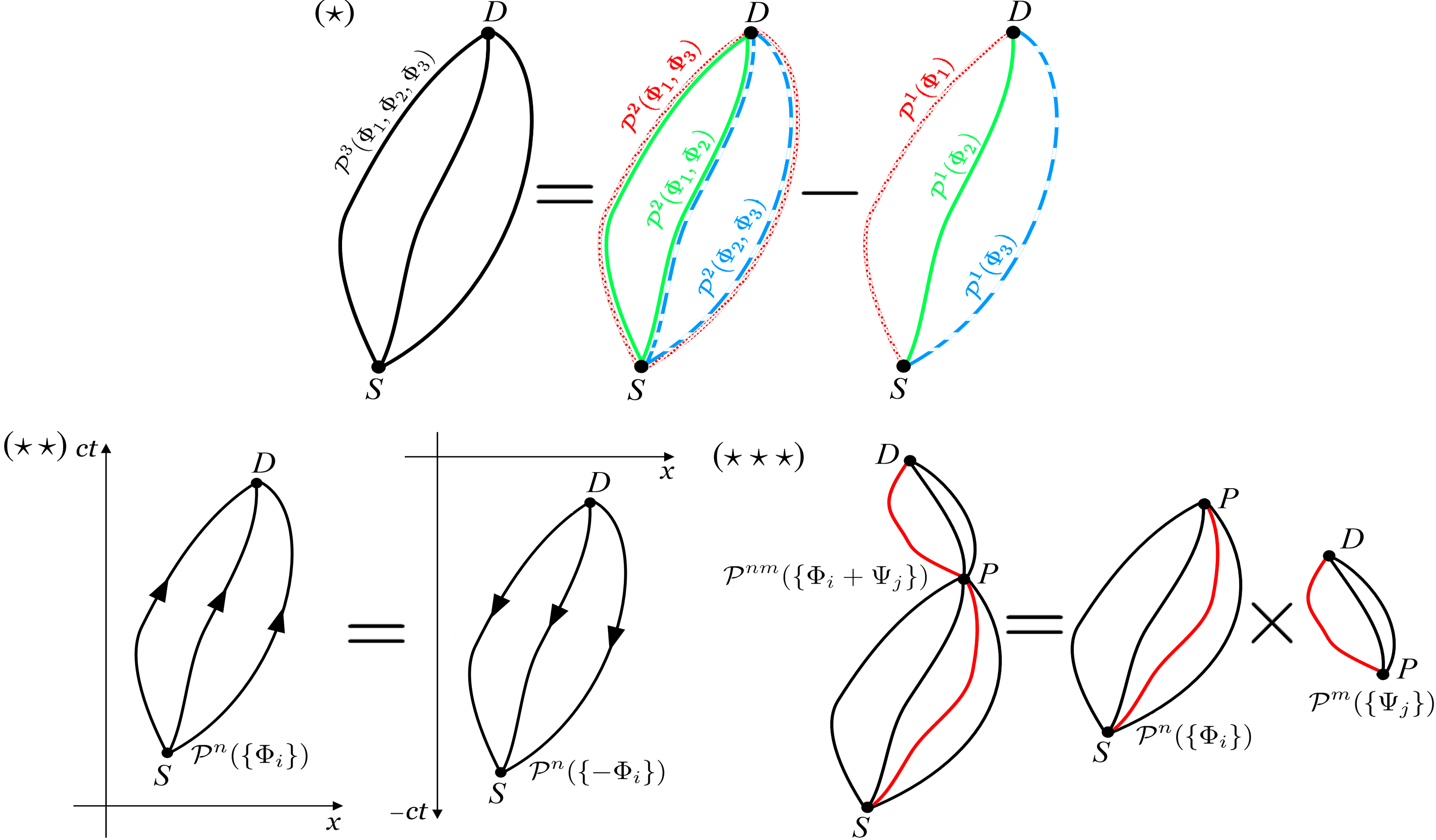}
    \caption{\justifying Illustration of the three conditions imposed on the relativistically invariant probability functions $\mathcal{P}^n$: 
    ($\star$) \emph{Pairwise Kolmogorov additivity:} The probability for $n$ alternatives is completely determined by the one-path and two-path contributions, with no irreducible higher-order overlaps.
    ($\star\,\star$) \emph{Time symmetry:} $\mathcal{P}^n$ remains invariant under time reversal, so the direction of time between two spacetime points does not affect the probabilities. 
    ($\star\star\star$) \emph{Bayesian composition:} The probability for a transition from $S$ to $D$ via an intermediate point $P$ factorizes into the probabilities for $S \to P$ and $P \to D$, in analogy with the Bayesian chain rule. For any trajectory (exemplified by the red path), the proper time from $S$ to $D$ via $P$ is the sum of the proper times along the segments $S\to P$ and $P\to D$, since the spacetime-invariant line element is additive under path concatenation.}
    \label{fig:conditions}
\end{figure*}

\emph{Time symmetry} requires invariance of the probability functional under time reversal of all paths:
\begin{align} \label{eq:time-symmetry}
    \mathcal{P}^n(\{\Phi_i\}) = \mathcal{P}^n(\{-\Phi_i\}). \tag{$\star\,\star$}
\end{align}

\emph{Bayesian composition} expresses the probabilistic analogue of the Bayesian chain rule. The functional $\mathcal{P}^{nm}$ describes a two-stage process with $n$ paths from the source at $S$ to some intermediate point $P$ and $m$ paths from $P$ to the detector $D$, resulting in $nm$ composite paths from $S$ to $D$ via $P$:
\begin{align} \label{eq:bayes-rule}
    \mathcal{P}^{nm}(\{\Phi_i+\Psi_j\})
    = \mathcal{P}^{n}(\{\Phi_i\})\, \mathcal{P}^{m}(\{\Psi_j\}), \tag{$\star\star\star$}
\end{align}
where $\{\Phi_i\}$ and $\{\Psi_j\}$ denote the invariants along the $S\to P$ and $P\to D$ segments, respectively, while $\{\Phi_i+\Psi_j\}\equiv(\Phi_1+\Psi_1, \Phi_1 + \Psi_2,\ldots,\Phi_n+\Psi_m)$. This condition is a consistency requirement with classical probability composition, i.e.\ with the Chapman--Kolmogorov equation for a single discrete intermediate state~\cite{ross2014introduction}. These three structural conditions are illustrated in Fig.~\ref{fig:conditions}.

Our main claim is that the only bounded, smooth, and nontrivial family of probability functions $\mathcal{P}^n$ satisfying conditions~\eqref{eq:pairwise-decomposition}, \eqref{eq:time-symmetry}, and~\eqref{eq:bayes-rule} is given by
\begin{align} \label{eq:final-claim}
    \mathcal{P}^n(\{\Phi_i\})
    = \left| \sum_{i=1}^n \ee^{\ii \kappa \Phi_i} \right|^{2},
\end{align}
where $\kappa$ is an arbitrary real constant with dimensions inverse to those of the chosen relativistic invariant $\Phi$ (proper time, action, etc.). In this form $\mathcal{P}^n$ should be interpreted as an unnormalized probability density. As in the case of the Feynman path integrals~\cite{feynman1948space,FeynmanHibbs1965}, the corresponding functional integration measure is subtle and often only defined as a suitable limit of discrete approximations. A mathematically rigorous reconstruction of path integrals from discrete approximations, including the nontrivial issue of normalization of the corresponding measures, is discussed in detail in Ref.~\cite{Albeverio2008PathIntegrals}.

Choosing $\kappa = \frac{1}{\hbar}$ and $\Phi_i = -S_i$ reproduces the Feynman path-integral rule \cite{feynman1948space,FeynmanHibbs1965}, showing that the prescription of adding complex amplitudes before squaring is not an arbitrary feature of quantum theory, but a mathematical consequence of Bayes' rule and relativistic invariance combined with the absence of interference beyond second order.

\paragraph*{Proof outline.} The derivation proceeds in four steps. (i) Setting $n=m=1$ in the Bayesian composition rule gives Cauchy's multiplicative equation, so $\mathcal{P}^1$ must be exponential; imposing time symmetry forces $\mathcal{P}^1\equiv 1$. (ii) Using time symmetry and the $m=1$ case of Bayesian composition, $\mathcal{P}^2(\Phi_1,\Phi_2)$ reduces to an even single-variable function $\mathcal{D}(\Phi_1-\Phi_2)$. (iii) Pairwise Kolmogorov additivity then expresses every $\mathcal{P}^n$ as a symmetric sum over pairwise differences in terms of $\mathcal{D}$, reducing the entire family to determining $\mathcal{D}$. (iv) Evaluating Bayesian composition for $n=m=2$ yields a closed functional equation for $\mathcal{D}$, reducible by $\mathcal{D}(x)=2G(x)+2$ to d'Alembert's equation; continuity gives $G\equiv 0$, $G(x)=\cos(\kappa x)$, or $G(x)=\cosh(\lambda x)$, and boundedness excludes the hyperbolic branch. Substituting back yields $\mathcal{P}^n(\{\Phi_i\})=\left|\sum_{i=1}^n e^{\ii\kappa\Phi_i}\right|^2$ as the unique bounded nontrivial solution.

\section{Proof}

\paragraph*{One-argument Bayes rule.}

We seek the most general family of probability functions $ \mathcal{P}^n $ satisfying the stated conditions. In the simplest instance of Eq.~\eqref{eq:bayes-rule} with $ n=m=1 $, the Bayesian composition law reduces to a functional equation for a single real argument. In particular, we obtain Cauchy’s multiplicative functional equation \cite{aczel2006lectures},
\begin{align}  \label{eq:bayes-rule-cauchy}
    \mathcal{P}^{1}(\Phi+\Psi) = \mathcal{P}^{1}(\Phi)\,\mathcal{P}^{1}(\Psi),
\end{align}
whose (continuous) solutions are exponential:
\begin{align}  \label{eq:cauchy-solution}
    \mathcal{P}^{1}(\Phi) = \exp\!\big[ \partial\mathcal{P}^{1}(0)\,\Phi \big],
\end{align}
where $\partial\mathcal{P}^{1}(0)$ denotes the derivative of $\mathcal{P}^{1}$ at $0$. This identifies the elementary building block: under composition of segments (addition of invariants) the one-argument function multiplies, hence it must be exponential in its argument.

\paragraph*{\texorpdfstring{$n$}{n}-argument Bayes rule.}

We now turn to the many-argument case and consider Eq.~\eqref{eq:bayes-rule} with arbitrary $n$ and $m=1$:
\begin{align}  \label{eq:bayes-rule-1}
    \mathcal{P}^{n}(\{\Phi_i + \Psi\})
    = \mathcal{P}^{n}(\{\Phi_i\}) \, \mathcal{P}^{1}(\Psi).
\end{align}
This identity expresses the invariance of $\mathcal{P}^{n}$ under a uniform shift of all arguments by the same increment $\Psi$, compensated by the one-argument factor.

Let us denote $\overline{\Phi} \equiv \frac{1}{n} \sum_{i=1}^n \Phi_i$, consider a special case of $\Psi = - \overline{\Phi}$, plug it into into Eq.~\eqref{eq:bayes-rule-1}, use Eq.~\eqref{eq:cauchy-solution} and carry out some basic rearrangements:
\begin{align}
    \mathcal{P}^{n}\big(\big\{\Phi_i-\overline \Phi\big\}\big) & = \mathcal{P}^n(\{\Phi_i\}) \mathcal{P}^{1}(-\overline{\Phi}),\\
    \mathcal{P}^n(\{\Phi_i\}) &= \mathcal{P}^{n}\big(\big\{\Phi_i-\overline \Phi\big\}\big) \exp\!\left[\partial\mathcal{P}^{1}(0)\,\overline{\Phi}\right]. \label{eq:R-solution}
\end{align}
We see that $\mathcal{P}^n$ factorizes into an $n$-argument function of the deviations of its arguments from their arithmetic mean, and an exponential function of the arithmetic mean itself.

\paragraph*{Time symmetry.}

We now impose time-reversal symmetry. For $n=1$, condition~\eqref{eq:time-symmetry} together with Eq.~\eqref{eq:cauchy-solution} implies
\begin{align} 
    &\forall_\Phi \; \exp\!\big[ \partial\mathcal{P}^{1}(0)\,\Phi \big]  = \exp\!\big[ -\partial\mathcal{P}^{1}(0)\,\Phi \big] \nonumber\\
    &\Rightarrow \quad \partial\mathcal{P}^{1}(0) = 0 
    \quad \Rightarrow \quad 
    \mathcal{P}^{1}(\Phi)=1. \label{eq:P10}
\end{align}
Moreover, the functional $\mathcal{P}^n(\{\Phi_i\})$ must satisfy \eqref{eq:time-symmetry}, and with Eq.~\eqref{eq:P10}, Eq.~\eqref{eq:R-solution} yields
\begin{align}
    \mathcal{P}^n(\{\Phi_i\}) = \mathcal{P}^{n}\left(\left\{\Phi_i - \overline{\Phi} \right\} \right)= \mathcal{P}^{n}\left(\left\{-\Phi_i + \overline{\Phi} \right\} \right).
\end{align} 
In particular, for $n=2$ the two-argument functional can be written directly as
\begin{align} \label{eq:P2-D2}
    \mathcal{P}^{2}(\Phi_1,\Phi_2) &=\mathcal{P}^{2}\left(\frac{\Phi_1-\Phi_2}{2},-\frac{\Phi_1-\Phi_2}{2}\right) \nonumber \\
    &\equiv \mathcal{D}(\Phi_1 - \Phi_2),
\end{align}
with $\mathcal{D}$ an even function of a single real variable.

\paragraph*{Pairwise Kolmogorov additivity.}

We now combine the pairwise Kolmogorov condition~\eqref{eq:pairwise-decomposition} with Eq.~\eqref{eq:P10} and Eq.~\eqref{eq:P2-D2}. For any $n \ge 2$, this yields
\begin{align} \label{eq:pn-D2}
\mathcal{P}^{n}(\{\Phi_i\}) &=   \sum_{1 \le i < j \le n} \mathcal{P}^2(\Phi_i,\Phi_j)
    - (n-2)\sum_{i=1}^n \mathcal{P}^1(\Phi_i) \nonumber\\
&= \sum_{1 \le i < j \le n} \mathcal{D}(\Phi_i - \Phi_j) - (n - 2)n,
\end{align}
so that all $\mathcal{P}^n$ are expressed in terms of a single even function~$\mathcal{D}$.

\paragraph*{\texorpdfstring{$nm$}{nm}-argument Bayes rule.}

To determine the explicit form of $\mathcal{D}$, we evaluate Eq.~\eqref{eq:bayes-rule} for $n=m=2$,
\begin{align}\label{eq:bayes-rule-2x2}
    \mathcal{P}^4(&\Phi_1+\Psi_1,\ \Phi_1+\Psi_2,\ \Phi_2+\Psi_1,\ \Phi_2+\Psi_2) \nonumber \\
    &= \mathcal{P}^2(\Phi_1,\Phi_2)\,\mathcal{P}^2(\Psi_1,\Psi_2).
\end{align}
Using Eq.~\eqref{eq:pn-D2} on the left (with $n=4$) and Eq.~\eqref{eq:P2-D2} on the right, set $A_1:=\Phi_1+\Psi_1, A_2:=\Phi_1+\Psi_2, A_3:=\Phi_2+\Psi_1, A_4:=\Phi_2+\Psi_2$. Then Eq.~\eqref{eq:pn-D2} gives
\begin{align}\label{eq:P4-pairsum}
   \mathcal{P}^4(A_1,A_2,A_3,A_4) =\sum_{1\le i<j\le 4}\mathcal{D}(A_i-A_j)-8.
\end{align}
Introducing $x:=\Phi_1-\Phi_2$ and $y:=\Psi_1-\Psi_2$, the pairwise differences are $A_1-A_2=y, A_3-A_4=y, A_1-A_3=x, A_2-A_4=x, A_1-A_4=x+y, A_2-A_3=x-y$, hence
\begin{align}\label{eq:P4-expanded}
   \mathcal{P}^4(A_1,A_2,A_3,A_4) =&\mathcal{D}(x+y)+\mathcal{D}(x-y) \nonumber\\
    &+2\mathcal{D}(x)+2\mathcal{D}(y)-8.
\end{align}
Moreover, by Eq.~\eqref{eq:P2-D2}, $\mathcal{P}^2(\Phi_1,\Phi_2)=\mathcal{D}(x)$ and $\mathcal{P}^2(\Psi_1,\Psi_2)=\mathcal{D}(y)$. Substituting into Eq.~\eqref{eq:bayes-rule-2x2} yields the functional equation
\begin{align}\label{eq:functional-equation-G}
    \mathcal{D}(x+y)+\mathcal{D}(x-y)+2\mathcal{D}(x)+2\mathcal{D}(y)-8 =\mathcal{D}(x)\mathcal{D}(y).
\end{align}

With $\mathcal{D}(x)=2G(x)+2$, Eq.~\eqref{eq:functional-equation-G} reduces to the d'Alembert equation
\begin{align}\label{eq:dalembert}
    G(x+y)+G(x-y)=2G(x)G(y),
\end{align}
whose continuous solutions are $G\equiv 0$, $G(x)=\cos(\kappa x)$, or $G(x)=\cosh(\lambda x)$ \cite{aczel2006lectures}. As we want our theory to describe bounded, nontrivial probability functions, the trivial solution and the unbounded hyperbolic branch have to be discarded, hence
\begin{align}\label{eq:D-cos}
    \mathcal{D}(x)=2\cos(\kappa x)+2.
\end{align}
Substituting Eq.~\eqref{eq:D-cos} into Eq.~\eqref{eq:pn-D2} gives
\begin{align}
    \mathcal{P}^{n}(\{\Phi_i\}) &= \sum_{1\le i<j\le n}\Big[2\cos\bigl(\kappa(\Phi_i-\Phi_j)\bigr)+2\Big]-n(n-2)\nonumber\\
    &=\left|\sum_{i=1}^n e^{\ii\kappa\Phi_i}\right|^2,
\end{align}
which is the most general bounded, nontrivial solution consistent with Eqs.~\eqref{eq:pairwise-decomposition}, \eqref{eq:time-symmetry}, and \eqref{eq:bayes-rule}.

The Kolmogorov condition \eqref{eq:pairwise-decomposition} strongly constrains the possible form of $\mathcal{P}^n$, imposing a restricted pairwise additive structure. Condition~\eqref{eq:bayes-rule} enforces a multiplicative composition law --- leading, together with condition~\eqref{eq:time-symmetry}, to oscillatory or hyperbolic solutions. The choice of trigonometric solutions arises from the normalizability of the probabilistic expression in the limit of infinitely many paths. Together they yield the squared sum of complex exponentials, showing that the quantum rule of amplitude addition is not arbitrary but follows directly from relativistic invariance and the elementary properties of a minimally nonclassical probabilistic theory.

We may also consider more general probabilistic theories in which condition~\eqref{eq:pairwise-decomposition} is not imposed. Such theories are then required to satisfy only conditions~\eqref{eq:time-symmetry} and~\eqref{eq:bayes-rule}. Appendix~\ref{app:fourier-solution} provides a closed-form Fourier-space solution, which accommodates this generalization.\\

\section{Discussion}
The presented derivation is a contribution to the relativistic programme initiated by Dragan and Ekert \cite{DraganEkert2020}. Their key physical input is that allowing the superluminal branch of inertial transformations motivates an effective multi-path description, so that relativistically invariant probabilities must be assigned to sets of alternative trajectories and may depend only on additive path invariants (e.g. proper time/action). At the structural level, \cite{DraganEkert2020} imposes Bayesian (chain-rule) composition for successive segments and time-reversal symmetry, which points toward oscillatory (amplitude-like) behavior but does not yield a uniqueness theorem. We close the reconstruction by adding a third, operationally motivated input: the Sorkin-inspired absence of irreducible higher-than-second-order interference \cite{sorkin1994quantum,sorkin2007quantum}, implemented as Kolmogorov-style \emph{pairwise additivity}. This fixes $\mathcal{P}^n$ in terms of $\mathcal{P}^1$ and $\mathcal{P}^2$ and collapses the remaining functional freedom to a single bounded oscillatory family.

Two points sharpen the comparison with \cite{DraganEkert2020,Dragan2022}. First, explicit permutation invariance of $\mathcal{P}^n$ (independence from path indexing) is not an independent axiom once conditions ~\eqref{eq:pairwise-decomposition}, \eqref{eq:time-symmetry}, \eqref{eq:bayes-rule} are imposed: the reduction of $\mathcal{P}^n$ to symmetric sums over pairwise differences makes relabelling invariance automatic. Second, our appendix quantifies how large the solution space remains under the Dragan--Ekert conditions alone (Bayesian composition and time symmetry): it contains infinitely many functional degrees of freedom (including independent functional parameters associated with prime numbers of paths), so the pairwise-additivity constraint is exactly what converts the relativistic motivation into a closed reconstruction.

Within the framework of relativistically invariant probability functions depending on path invariants $\{\Phi_i\}$, we have identified three natural requirements — pairwise Kolmogorov additivity (equivalently, the absence of higher-order interference), time-reversal symmetry, and Bayesian composition of successive events — that uniquely lead to a nontrivial solution: the squared modulus of a sum of complex exponentials of these invariants. Rather than introducing the Hilbert-space structure \cite{von2018mathematical,mackey2004mathematical,gleason1975measures,birkhoff1975logic,piron1976foundations,soler1995characterization,sorkin1994quantum,sorkin2007quantum} or adopting information-theoretic \cite{Wheeler1989,hardy2001quantum,hardy2011reformulating,fuchs2002quantum,zeilinger1999foundational,brukner2003information,brukner2009information,grinbaum2003elements,chiribella2011informational,chiribella2016quantum,wootters1981statistical,BohrUlfbeck95,rovelli1996relational,clifton2003characterizing,DAriano08,masanes2011derivation,muller2012structure} axioms at the outset, the complex-amplitude framework emerges directly from minimal probabilistic assumptions together with relativistic invariance. Our result shows that the familiar Feynman prescription of adding complex amplitudes and then squaring is not an \emph{ad hoc} rule, but the natural consequence of combining relativistic invariance with basic probabilistic consistency.

\section*{Acknowledgments}
We acknowledge a very useful observation made by ChatGPT-5 that helped us complete the proof presented in this work. K.D. acknowledges project CZ.02.01.01/00/22\_008/0004649 (QUEENTEC) of the EU and the MEYS of the Czech Republic, and the project No. LUC25006 of MEYS, Czech Republic.

\bibliography{refs}

@book{Feynman1965LecturesIII,
    title={The Feynmann Lectures on Physics, Vol.~III: Quantum Mechanics},
    author={Feynman, Richard P. and Leighton, Robert B. and Sands, Matthew},
    year={1963},
    publisher={Addison-Wesley, California Institute of Technology},
    url={https://www.feynmanlectures.caltech.edu/III_01.html}
}

@article{DraganEkert2020,
    author = {Dragan, Andrzej and Ekert, Artur},
    title     = {Quantum Principle of Relativity},
    journal     = {New Journal of Physics},
    year        = {2020},
    volume    = {22},
    pages     = {033038},
    doi         = {10.1088/1367-2630/ab76f7}
}

@article{Dragan2022,
    author    = {Dragan, Andrzej and Dębski, Kacper and Charzyński, Szymon and Turzyński, Krzysztof and Ekert, Artur},
    title     = {Relativity of superluminal observers in 1+3 spacetime},
    journal     = {Classical and Quantum Gravity},
    year        = {2022},
    volume    = {40},
    pages     = {025013},
    doi         = {10.1088/1361-6382/acad60}
}

@book{Kolmogorov1950,
    author    = {Kolmogorov, Andrey Nikolaevich},
    title     = {Foundations of the Theory of Probability},
    year        = {1950},
    publisher = {Chelsea Publishing Company},
    address     = {New York},
    note        = {Translated from the original German edition published in 1933},
    url = {https://archive.org/details/foundationsofthe00kolm}
}

@article{sorkin1994quantum,
    title={Quantum mechanics as quantum measure theory},
    author={Sorkin, Rafael D},
    journal={Modern Physics Letters A},
    volume={9},
    number={33},
    pages={3119--3127},
    year={1994},
    publisher={World Scientific},
    doi={10.1142/S021773239400294X}
}

@article{sorkin2007quantum,
    title={Quantum dynamics without the wavefunction},
    author={Sorkin, Rafael D},
    journal={Journal of Physics A: Mathematical and Theoretical},
    volume={40},
    number={12},
    pages={3207},
    year={2007},
    publisher={IOP Publishing},
    doi={10.1088/1751-8113/40/12/S20}
}

@book{aczel2006lectures,
  title={Lectures on Functional Equations and Their Applications},
  author={Aczel, J. and Oser, H.},
  isbn={9780486445236},
  lccn={2005053735},
  series={Dover Books on Mathematics},
  url={https://books.google.pl/books?id=vPmJAwAAQBAJ},
  year={2006},
  publisher={Dover Publications}
}

@book{Dragan2021,
  title={Unusually Special Relativity},
  author={Dragan, A.},
  isbn={9781800610828},
  lccn={2021026275},
  url={https://books.google.pl/books?id=BIdYEAAAQBAJ},
  year={2021},
  publisher={World Scientific Publishing Company}
}

@article{sinha2010ruling,
  title={Ruling out multi-order interference in quantum mechanics},
  author={Sinha, Urbasi and Couteau, Christophe and Jennewein, Thomas and Laflamme, Raymond and Weihs, Gregor},
  journal={Science},
  volume={329},
  number={5990},
  pages={418--421},
  year={2010},
  publisher={American Association for the Advancement of Science},
  doi={10.1126/science.1190545}
}

@article{kauten2017obtaining,
  title={Obtaining tight bounds on higher-order interferences with a 5-path interferometer},
  author={Kauten, Thomas and Keil, Robert and Kaufmann, Thomas and Pressl, Benedikt and Brukner, {\v{C}}aslav and Weihs, Gregor},
  journal={New Journal of Physics},
  volume={19},
  number={3},
  pages={033017},
  year={2017},
  publisher={IOP Publishing},
  doi={10.1088/1367-2630/aa5d98}
}

@book{ross2014introduction,
  title={Introduction to Probability Models},
  author={Ross, S.M.},
  isbn={9780124081215},
  lccn={2013035819},
  url={https://books.google.pl/books?id=A3YpAgAAQBAJ},
  year={2014},
  publisher={Academic Press}
}

@article{feynman1948space,
  title={Space-time approach to non-relativistic quantum mechanics},
  author={Feynman, Richard Phillips},
  journal={Reviews of modern physics},
  volume={20},
  number={2},
  pages={367},
  year={1948},
  publisher={APS},
  doi={10.1103/RevModPhys.20.367}
}

@book{FeynmanHibbs1965,
    author    = {Richard P. Feynman and Albert R. Hibbs},
    title     = {Quantum Mechanics and Path Integrals},
    publisher = {McGraw–Hill},
    address     = {New York},
    year        = {1965},
    url={https://archive.org/details/quantummechanics0000feyn}
}

@book{Albeverio2008PathIntegrals,
  title={Mathematical theory of Feynman path integrals: an introduction},
  author={Albeverio, Sergio A and H{\o}egh-Krohn, Raphael J and Mazzucchi, Sonia},
  title     = {Mathematical Theory of Feynman Path Integrals},
  series    = {Lecture Notes in Mathematics},
  year={2008},
  publisher={Springer},
  address     = {Berlin},
  doi={10.1007/978-3-540-76956-9}
}

@article{goyal2010origin,
  title={Origin of complex quantum amplitudes and Feynman’s rules},
  author={Goyal, Philip and Knuth, Kevin H and Skilling, John},
  journal={Physical Review A},
  volume={81},
  number={2},
  pages={022109},
  year={2010},
  publisher={APS},
  doi={10.1103/PhysRevA.81.022109}
}

@article{knuth2012foundations,
  title={Foundations of inference},
  author={Knuth, Kevin H and Skilling, John},
  journal={Axioms},
  volume={1},
  number={1},
  pages={38--73},
  year={2012},
  publisher={Molecular Diversity Preservation International},
  doi={10.3390/axioms1010038}
}

@article{hardy2001quantum,
      title={Quantum Theory From Five Reasonable Axioms}, 
      author={Lucien Hardy},
      year={2001},
      journal={arXiv:quant-ph/0101012},
      url={https://arxiv.org/abs/quant-ph/0101012}
}

@article{hardy2011reformulating,
  title={Reformulating and reconstructing quantum theory},
  author={Hardy, Lucien},
  journal={arXiv:1104.2066},
  year={2011},
  url={https://doi.org/10.48550/arXiv.1104.2066}
}

@article{chiribella2011informational,
  title={Informational derivation of quantum theory},
  author={Chiribella, Giulio and D’Ariano, Giacomo Mauro and Perinotti, Paolo},
  journal={Physical Review A},
  volume={84},
  number={1},
  pages={012311},
  year={2011},
  publisher={APS},
  doi={10.1103/PhysRevA.84.012311}
}

@book{chiribella2016quantum,
  title={Quantum theory: informational foundations and foils},
  author={Chiribella, Giulio and Spekkens, Robert W},
  year={2016},
  publisher={Springer},
  do={10.1007/978-94-017-7303-4}
}

@article{BohrUlfbeck95,
  title={Primary manifestation of symmetry. Origin of quantal indeterminacy},
  author={Bohr, Aage and Ulfbeck, Ole},
  journal={Reviews of Modern Physics},
  volume={67},
  number={1},
  pages={1},
  year={1995},
  publisher={APS},
  doi={10.1103/RevModPhys.67.1}
}

@article{zeilinger1999foundational,
  title={A foundational principle for quantum mechanics},
  author={Zeilinger, Anton},
  journal={Foundations of Physics},
  volume={29},
  number={4},
  pages={631--643},
  year={1999},
  publisher={Springer},
  doi={10.1023/A:1018820410908}
}

@incollection{brukner2003information,
  title={Information and fundamental elements of the structure of quantum theory},
  author={Brukner, {\v{C}}aslav and Zeilinger, Anton},
  booktitle={Time, quantum and information},
  pages={323--354},
  year={2003},
  publisher={Springer},
  doi={10.1007/978-3-662-10557-3_21}
}

@article{brukner2009information,
  title={Information invariance and quantum probabilities},
  author={Brukner, {\v{C}}aslav and Zeilinger, Anton},
  journal={Foundations of Physics},
  volume={39},
  number={7},
  pages={677--689},
  year={2009},
  publisher={Springer},
  doi={10.1007/s10701-009-9316-7}
}

@article{grinbaum2003elements,
  title={Elements of information-theoretic derivation of the formalism of quantum theory},
  author={Grinbaum, Alexei},
  journal={International Journal of Quantum Information},
  volume={1},
  number={03},
  pages={289--300},
  year={2003},
  publisher={World Scientific},
  doi={10.1142/S0219749903000309}
}

@article{wootters1981statistical,
  title={Statistical distance and Hilbert space},
  author={Wootters, William K},
  journal={Physical Review D},
  volume={23},
  number={2},
  pages={357},
  year={1981},
  publisher={APS},
  doi={10.1103/PhysRevD.23.357}
}

@incollection{DAriano08,
  author    = {D'Ariano, Giacomo Mauro},
  title     = {Probabilistic theories: what is special about Quantum Mechanics?},
  booktitle = {Philosophy of Quantum Information and Entanglement},
  editor    = {Bokulich, Alisa and Jaeger, Gordon},
  publisher = {Cambridge University Press},
  isbn      ={9781139487665},
  year      = {2010},
  doi       = {10.48550/arXiv.0807.4383}
}

@article{rovelli1996relational,
  title={Relational quantum mechanics},
  author={Rovelli, Carlo},
  journal={International journal of theoretical physics},
  volume={35},
  number={8},
  pages={1637--1678},
  year={1996},
  publisher={Springer},
  doi={10.1007/BF02302261}
}

@article{clifton2003characterizing,
  title={Characterizing quantum theory in terms of information-theoretic constraints},
  author={Clifton, Rob and Bub, Jeffrey and Halvorson, Hans},
  journal={Foundations of Physics},
  volume={33},
  number={11},
  pages={1561--1591},
  year={2003},
  publisher={Springer},
  doi={10.1023/A:1026056716397}
}

@incollection{birkhoff1975logic,
  title={The logic of quantum mechanics},
  author={Birkhoff, Garrett and Von Neumann, John},
  booktitle={The Logico-Algebraic Approach to Quantum Mechanics: Volume I: Historical Evolution},
  pages={1--26},
  year={1975},
  publisher={Springer},
  doi={10.1007/978-94-010-1795-4_1}
}

@book{mackey2004mathematical,
  title={Mathematical foundations of quantum mechanics},
  author={Mackey, George W},
  year={2004},
  publisher={Courier Corporation},
  url={https://books.google.pl/books?id=XN4oAwAAQBAJ}
}

@incollection{piron1976foundations,
  title={On the foundations of quantum physics},
  author={Piron, Constantin},
  booktitle={Quantum Mechanics, Determinism, Causality, and Particles: An International Collection of Contributions in Honor of Louis de Broglie on the Occasion of the Jubilee of His Celebrated Thesis},
  pages={105--116},
  year={1976},
  publisher={Springer},
  doi={10.1007/978-94-010-1440-3_7}
}

@incollection{gleason1975measures,
  title={Measures on the closed subspaces of a Hilbert space},
  author={Gleason, Andrew M},
  booktitle={The Logico-Algebraic Approach to Quantum Mechanics: Volume I: Historical Evolution},
  pages={123--133},
  year={1975},
  publisher={Springer},
  doi={10.1007/978-94-010-1795-4_7}
}

@article{soler1995characterization,
  title={Characterization of Hilbert spaces by orthomodular spaces},
  author={Sol{\`e}r, Maria Pia},
  journal={Communications in Algebra},
  volume={23},
  number={1},
  pages={219--243},
  year={1995},
  publisher={Taylor \& Francis},
  doi={10.1080/00927879508825218}
}

@article{cox1946probability,
  title={Probability, frequency and reasonable expectation},
  author={Cox, Richard T},
  journal={American journal of physics},
  volume={14},
  number={1},
  pages={1--13},
  year={1946},
  publisher={American Association of Physics Teachers},
  doi={10.2307/2272983}
}

@book{cox2001algebra,
  title={Algebra of probable inference},
  author={Cox, Richard T},
  year={2001},
  publisher={Johns Hopkins University Press},
  doi={10.56021/9780801869822}
}

@incollection{Wheeler1989,
	author = {John Archibald Wheeler},
	booktitle = {Proceedings III International Symposium on Foundations of Quantum Mechanics},
	editor = {Wheeler John Archibald},
	pages = {354--368},
	title = {Information, Physics, Quantum: The Search for Links},
	year = {1989},
    publisher={Physical Society of Japan},
    url={https://philpapers.org/rec/WHEIPQ}
}

@book{von2018mathematical,
  title={Mathematical foundations of quantum mechanics: New edition},
  author={Von Neumann, John},
  year={2018},
  publisher={Princeton university press},
  url={https://books.google.pl/books?id=B3OYDwAAQBAJ}
}

@article{masanes2011derivation,
  title={A derivation of quantum theory from physical requirements},
  author={Masanes, Llu{\'\i}s and M{\"u}ller, Markus P},
  journal={New Journal of Physics},
  volume={13},
  number={6},
  pages={063001},
  year={2011},
  publisher={IOP Publishing},
  doi={10.1088/1367-2630/13/6/063001}
}

@article{popescu1994quantum,
  title={Quantum nonlocality as an axiom},
  author={Popescu, Sandu and Rohrlich, Daniel},
  journal={Foundations of Physics},
  volume={24},
  number={3},
  pages={379--385},
  year={1994},
  publisher={Springer},
  doi={10.1007/BF02058098}
}

@articel{dam2005,
  title={Implausible Consequences of Superstrong Nonlocality},
  author={Wim van Dam},
  year={2005},
  journal={arXiv:quant-ph/0501159},
  url={https://doi.org/10.48550/arXiv.quant-ph/0501159}
}

@article{brassard2006limit,
  title={Limit on nonlocality in any world in which communication complexity is not trivial},
  author={Brassard, Gilles and Buhrman, Harry and Linden, Noah and M{\'e}thot, Andr{\'e} Allan and Tapp, Alain and Unger, Falk},
  journal={Physical Review Letters},
  volume={96},
  number={25},
  pages={250401},
  year={2006},
  publisher={APS},
  doi={10.1103/PhysRevLett.96.250401}
}

@article{barnum2007generalized,
  title={Generalized no-broadcasting theorem},
  author={Barnum, Howard and Barrett, Jonathan and Leifer, Matthew and Wilce, Alexander},
  journal={Physical review letters},
  volume={99},
  number={24},
  pages={240501},
  year={2007},
  publisher={APS},
  doi={10.1103/PhysRevLett.99.240501}
}

@article{fuchs2002quantum,
  title={Quantum mechanics as quantum information (and only a little more)},
  author={Fuchs, Christopher A},
  journal={arXiv:quant-ph/0205039},
  year={2002},
  url={https://doi.org/10.48550/arXiv.quant-ph/0205039}
}

@article{pawlowski2009information,
  title={Information causality as a physical principle},
  author={Paw{\l}owski, Marcin and Paterek, Tomasz and Kaszlikowski, Dagomir and Scarani, Valerio and Winter, Andreas and {\.Z}ukowski, Marek},
  journal={Nature},
  volume={461},
  number={7267},
  pages={1101--1104},
  year={2009},
  publisher={Nature Publishing Group UK London},
  doi={10.1038/nature08400}
}

@article{allcock2009recovering,
  title={Recovering part of the boundary between quantum and nonquantum correlations from information causality},
  author={Allcock, Jonathan and Brunner, Nicolas and Pawlowski, Marcin and Scarani, Valerio},
  journal={Physical Review A—Atomic, Molecular, and Optical Physics},
  volume={80},
  number={4},
  pages={040103},
  year={2009},
  publisher={APS},
  doi={10.1103/PhysRevA.80.040103}
}

@article{navascues2015almost,
  title={Almost quantum correlations},
  author={Navascu{\'e}s, Miguel and Guryanova, Yelena and Hoban, Matty J and Ac{\'\i}n, Antonio},
  journal={Nature communications},
  volume={6},
  number={1},
  pages={6288},
  year={2015},
  publisher={Nature Publishing Group UK London},
  doi={10.1038/ncomms7288}
}

@article{muller2012structure,
  title={Structure of reversible computation determines the self-duality of quantum theory},
  author={M{\"u}ller, Markus P and Ududec, Cozmin},
  journal={Physical review letters},
  volume={108},
  number={13},
  pages={130401},
  year={2012},
  publisher={APS},
  doi={10.1103/PhysRevLett.108.130401}
}

@article{goyal2014,
  title = {Derivation of quantum theory from Feynman's rules},
  author = {Goyal, Philip},
  journal = {Phys. Rev. A},
  volume = {89},
  issue = {3},
  pages = {032120},
  numpages = {12},
  year = {2014},
  month = {Mar},
  publisher = {American Physical Society},
  doi = {10.1103/PhysRevA.89.032120}
}

@article{goyal2008,
  title = {Information-geometric reconstruction of quantum theory},
  author = {Goyal, Philip},
  journal = {Phys. Rev. A},
  volume = {78},
  issue = {5},
  pages = {052120},
  numpages = {17},
  year = {2008},
  month = {Nov},
  publisher = {American Physical Society},
  doi = {10.1103/PhysRevA.78.052120}
}

@article{grudka2022comment,
  title={Comment on ‘Quantum principle of relativity’},
  author={Grudka, Andrzej and W{\'o}jcik, Antoni},
  journal={New Journal of Physics},
  volume={24},
  number={9},
  pages={098001},
  year={2022},
  publisher={IOP Publishing},
  doi={10.1088/1367-2630/ac924e}
}

@article{grudka2023superluminal,
  title={Superluminal observers do not explain quantum superpositions},
  author={Grudka, Andrzej and Stempin, Jȩdrzej and W{\'o}jcik, Jan and W{\'o}jcik, Antoni},
  journal={Physics Letters A},
  volume={487},
  pages={129127},
  year={2023},
  publisher={Elsevier},
  doi={10.1016/j.physleta.2023.129127}
}

@article{del2023comment,
  title={Comment on ‘Quantum principle of relativity’},
  author={Del Santo, Flavio and Horvat, Sebastian},
  journal={New Journal of Physics},
  volume={24},
  number={12},
  pages={128001},
  year={2023},
  publisher={IOP Publishing},
  doi={10.1088/1367-2630/acae3b}
}

@article{horodecki2023comment,
  title={Comment on ‘Quantum principle of relativity’},
  author={Horodecki, Ryszard},
  journal={New Journal of Physics},
  volume={25},
  number={12},
  pages={128001},
  year={2023},
  publisher={IOP Publishing},
  doi={10.1088/1367-2630/ad10ff}
}

@article{Dragan2022reply,
doi = {10.1088/1367-2630/ac90e3},
year = {2022},
publisher = {IOP Publishing},
volume = {24},
number = {9},
pages = {098002},
author = {Dragan, Andrzej and Ekert, Artur},
title = {Reply to Comment on ‘Quantum principle of relativity’},
journal = {New Journal of Physics}
}

@article{Dragan2023reply2,
doi = {10.1088/1367-2630/acec62},
year = {2023},
publisher = {IOP Publishing},
volume = {25},
number = {8},
pages = {088002},
author = {Dragan, Andrzej and Ekert, Artur},
title = {Reply to Comment on ‘Quantum principle of relativity’},
journal = {New Journal of Physics}
}

@article{dragan2023reply,
  title={Reply to the Comment on ‘Quantum principle of relativity’},
  author={Dragan, Andrzej and Ekert, Artur},
  journal={New Journal of Physics},
  volume={25},
  number={12},
  pages={128002},
  year={2023},
  publisher={IOP Publishing},
  doi={10.1088/1367-2630/ad100e}
}

@article{gavassino2022fundamental,
  title={Fundamental asymmetry between space and time in quantum field theory},
  author={Gavassino, L},
  journal={Classical and Quantum Gravity},
  volume={39},
  number={21},
  pages={215010},
  year={2022},
  publisher={IOP Publishing},
  doi={10.1088/1361-6382/ac9107}
}

@article{lake2025,
  title={The (1+ 3)-dimensional ‘quantum principle of relativity’is Einstein’s principle of relativity},
  author={Lake, Matthew J},
  journal={The European Physical Journal C},
  volume={85},
  number={1},
  pages={92},
  year={2025},
  publisher={Springer},
  doi={10.1140/epjc/s10052-024-13667-9}
}

@article{sen2025superluminal,
  title={Superluminal Quantum Reference Frames},
  author={Sen, Amrapali and Salzger, Matthias and others},
  year={2025},
  journal={arXiv:2506.11787},
  url={https://doi.org/10.48550/arXiv.2506.11787}
}

@article{Damski2025, 
  title={Lorentz Transformations in 1+1 Dimensional Spacetime: Mainly the Superluminal Case}, 
  author={Damski, B.S.}, 
  journal={Acta Physica Polonica A}, 
  volume={148},  
  number={1}, 
  pages={22},
  year={2025},
  doi={10.12693/APhysPolA.148.22}
}

@article{paczos2024covariant,
  title={Covariant quantum field theory of tachyons},
  author={Paczos, Jerzy and D{\k{e}}bski, Kacper and Cedrowski, Szymon and Charzy{\'n}ski, Szymon and Turzy{\'n}ski, Krzysztof and Ekert, Artur and Dragan, Andrzej},
  journal={Physical Review D},
  volume={110},
  number={1},
  pages={015006},
  year={2024},
  publisher={APS},
  doi={10.1103/PhysRevD.110.015006}
}

@article{jodlowski2024covariant,
  title={Covariant quantum field theory of tachyons is unphysical},
  author={Jod{\l}owski, Krzysztof},
  journal={Physical Review D},
  volume={110},
  number={11},
  pages={115042},
  year={2024},
  publisher={APS},
  doi={10.1103/PhysRevD.110.115042}
}

@article{michalski2025stories,
  title={Stories in the two-state vector formalism},
  author={Michalski, Patryk and Dragan, Andrzej},
  journal={Journal of Physics Communications},
  volume={9},
  number={11},
  pages={115003},
  year={2025},
  publisher={IOP Publishing},
  doi={10.1088/2399-6528/ae0b2c}
}

@article{bayes,
    author = {Bayes, Thomas},
    title = {LII. An essay towards solving a problem in the doctrine of chances. By the late Rev. Mr. Bayes, F. R. S. communicated by Mr. Price, in a letter to John Canton, A. M. F. R. S},
    journal = {Philosophical Transactions},
    number = {53},
    pages = {370-418},
    year = {1763},
    month = {12},
    doi = {10.1098/rstl.1763.0053}
}

\appendix

\section{Fourier-space solution of the Bayesian functional equation and the time-symmetry condition}
\label{app:fourier-solution}

For completeness, we derive here the most general solution of the $n$-argument Bayesian composition rule in Fourier space. We start from Eq.~\eqref{eq:bayes-rule} with arbitrary $n$ and $m=1$,
\begin{align}  \label{eq:app-bayes-rule-1}
    \mathcal{P}^{n}(\{\Phi_i+\Psi\}) = \mathcal{P}^{n}(\{\Phi_i\})\,\mathcal{P}^{1}(\Psi).
\end{align}
This relation states that a uniform shift of all arguments by the same increment $\Psi$ changes $\mathcal{P}^{n}$ only by an overall multiplicative factor given by the one-argument functional $\mathcal{P}^{1}$. 

To extract the differential form of this condition, we subtract $\mathcal{P}^{n}(\{\Phi_i\})$ from both sides, divide by $\Psi$, and take the limit $\Psi\to 0$. In this way we obtain the directional derivative of $\mathcal{P}^n$ along the vector $(1,1,\dots,1)$ in the argument space:
\begin{align}
\label{eq:bayes-rule-2}
\sum_{i=1}^n \partial_i\mathcal{P}^n(\{\Phi_i\})
&:= \lim_{\Psi \to 0} \frac{\mathcal{P}^n(\{\Phi_i + \Psi\}) - \mathcal{P}^n(\{\Phi_i\})}{\Psi} \nonumber\\
&= \partial\mathcal{P}^{1}(0)\,\mathcal{P}^n(\{\Phi_i\}).
\end{align}
Equation~\eqref{eq:bayes-rule-2} is a first-order linear partial differential equation: the infinitesimal change of $\mathcal{P}^n$ under a common shift of all arguments is proportional to the value of $\mathcal{P}^n$ itself.

To solve this constraint, it is convenient to work in Fourier space. Multiplying Eq.~\eqref{eq:bayes-rule-2} by $\exp\!\big[ \ii \sum_{i=1}^n \alpha_i \Phi_i \big]$ and integrating over all $\Phi_i \in \mathbb{R}$, we introduce the Fourier transform
\begin{align}
\widetilde{\mathcal{P}}^n(\{\alpha_i\})
:= \int_{\mathbb{R}^n} \frac{\dd^n \Phi}{(2\pi)^n}\,
\mathcal{P}^n(\{\Phi_i\}) \exp\!\Big[ \ii \sum_{i=1}^n \alpha_i \Phi_i \Big],
\end{align}
and find that $\widetilde{\mathcal{P}}^n$ must satisfy the algebraic constraint
\begin{align}
\label{eq:bayes-rule-3}
\Big( \ii \sum_{i=1}^n \alpha_i - \partial\mathcal{P}^{1}(0) \Big)\,
\widetilde{\mathcal{P}}^n(\{\alpha_i\}) = 0.
\end{align}
Thus $\widetilde{\mathcal{P}}^n$ can be nonzero only on the hyperplane in $\{\alpha_i\}$-space where the linear form in parentheses vanishes. Equivalently, we may write
\begin{align}
\label{eq:bayes-rule-4}
\widetilde{\mathcal{P}}^n(\{\alpha_i\})
= \widetilde{\mathcal{A}}^n(\{\alpha_i\})\,
\delta \left( \sum_{i=1}^n \alpha_i + \ii\partial\mathcal{P}^{1}(0) \right),
\end{align}
where $\widetilde{\mathcal{A}}^n$ is arbitrary (for instance, a tempered distribution) on that hyperplane.

Performing the inverse Fourier transform in Eq.~\eqref{eq:bayes-rule-4} yields the general solution of Eq.~\eqref{eq:app-bayes-rule-1}:
\begin{align}
\label{eq:solution-1}
\mathcal{P}^{n}(\{\Phi_i\})
&= \int_{\mathbb{R}^n} \dd^n\alpha\;
\widetilde{\mathcal{A}}^{n}(\{\alpha_i\})\,
\delta\left( \sum_{i=1}^n \alpha_i + \ii \partial\mathcal{P}^{1}(0) \right) \nonumber\\
&\quad\times \exp\left[ -\ii \sum_{i=1}^n \alpha_i \Phi_i \right],
\end{align}
which is the most general Fourier representation consistent with the Bayes rule. The delta function implements the translation invariance: only combinations of conjugate variables with zero (or purely imaginary shifted) total sum contribute, reflecting invariance of $\mathcal{P}^n$ under a common shift of all $\Phi_i$.

We now supplement the Bayesian condition with time-reversal symmetry, Eq.~\eqref{eq:time-symmetry}. From the one-argument solution~\eqref{eq:cauchy-solution}, invariance under $\Phi \mapsto -\Phi$ implies that $\mathcal{P}^{1}$ must be an even function, which forces $\mathcal{P}^{1}(\Phi)= 1$ and hence $\partial\mathcal{P}^{1}(0)=0$. Indeed, if $\partial\mathcal{P}^{1}(0) \neq 0$, the exponential law would not be symmetric under time reversal and the symmetry would be violated.

Substituting $\partial\mathcal{P}^{1}(0)=0$ into Eq.~\eqref{eq:solution-1} and then averaging the resulting expression over $\Phi_i \mapsto -\Phi_i$ leads to a manifestly even form,
\begin{align}
\label{eq:solution-2}
\mathcal{P}^{n}(\{\Phi_i\})
&= \int_{\mathbb{R}^n} \dd^n\alpha\;
\widetilde{\mathcal{A}}^{n}(\{\alpha_i\})\,
\delta\!\left( \sum_{i=1}^n \alpha_i \right) \nonumber\\
&\quad\times \cos\!\Big( \sum_{i=1}^n \alpha_i \Phi_i \Big).
\end{align}
Thus, once time-reversal symmetry is imposed, the admissible solutions are even functions of the arguments and satisfy a zero-sum constraint in the conjugate variables.

Without imposing condition \eqref{eq:pairwise-decomposition}, the resulting function is not explicitly invariant under permutations of its arguments — yet physically, the ordering of path labels is immaterial. We therefore introduce an additional requirement:
\begin{align} \label{eq:perm}
    \mathcal{P}^n(\{\Phi_i\}) = \mathcal{P}^n(\{\Phi_{\sigma(i)}\}), \tag{$\circ$}
\end{align}
where $\sigma \in S_n$, and $S_n$ is the symmetric group of all $n!$ permutations of an $n$-element set. 

With condition \eqref{eq:perm} imposed, Eq.~\eqref{eq:solution-2} takes the form
\begin{align}
\label{eq:solution-3}
\mathcal{P}^{n}(\{\Phi_i\})
&= \int_{\mathbb{R}^n} \frac{\dd^n\alpha}{n!}\;
\widetilde{\mathcal{A}}^{n}(\{\alpha_i\})\,
\delta\!\left( \sum_{i=1}^n \alpha_i \right) \nonumber\\
&\quad\times \sum_{\sigma \in S_n} 
\cos\!\Big( \sum_{i=1}^n \alpha_i \Phi_{\sigma(i)} \Big).
\end{align}
Equation~\eqref{eq:final-claim} is a particular solution of the general form obtained from conditions \eqref{eq:time-symmetry}, \eqref{eq:bayes-rule}, and \eqref{eq:perm}, and corresponds to the special case of Eq.~\eqref{eq:solution-3} that additionally satisfies the pairwise additivity constraint \eqref{eq:pairwise-decomposition}.

It is worth stressing that Bayesian composition in the multi-path setting is not a single functional equation but an \emph{infinite hierarchy}: for every pair $(n,m)$ it links an $nm$-argument functional to an $n$-argument and an $m$-argument one through the composition ${\Phi_i},{\Psi_j}\mapsto{\Phi_i+\Psi_j}$. Without an additional principle tying different arities together, this hierarchy enforces consistency under concatenation but does not close the reconstruction. One may still choose independent functional data at infinitely many arities—most transparently at each prime $p$—and the Bayes hierarchy then propagates these choices to composite $n$ via prime factorization. Hence the Dragan--Ekert conditions alone leave a generically infinite-dimensional solution space: the constraints relate different $(n,m)$ sectors, but they do not remove the freedom residing in the prime ones.

\vspace{-\parskip}
The pairwise-additivity assumption \eqref{eq:pairwise-decomposition} introduced in the main text supplies the missing \emph{closure across arity} absent from the conditions considered in \cite{DraganEkert2020}. The Kolmogorov/Sorkin decomposition expresses every $\mathcal{P}^n$ in terms of $\mathcal{P}^2$ (and $\mathcal{P}^1$), so higher-$n$ sectors carry no independent information and the prime-indexed freedoms disappear.

\end{document}